\documentclass[11pt,twoside]{article}

\usepackage{asp2006}

\begin{document}
\title{The Red-Sequence Cluster Surveys}
\author{H.K.C. Yee\altaffilmark{1}, M.D. Gladders\altaffilmark{2},
David G. Gilbank\altaffilmark{1}, S.~Majumdar\altaffilmark{3},
H. Hoekstra\altaffilmark{4}, E. Ellingson\altaffilmark{5},
and the RCS--2 Collaboration\altaffilmark{6}}   
\altaffiltext{1}{Dept.~of Astronomy \&~Astrophysics, Univ.~of Toronto,
Toronto, ON, M5S 3H4, Canada}
\altaffiltext{2}{Dept. of Astronomy \&~Astrophysics,
Univ.~of Chicago, Chicago, IL 60637, USA}
\altaffiltext{3}{CITA, Univ.~of Toronto,
Toronto, ON, M5S 3H8, Canada}
\altaffiltext{4}{Dept.~of Physics \&~Astronomy, Univ.~of Victoria,
Vitoria, BC, V8P 5C2, Canada}
\altaffiltext{5}{CASA, Univ.~of Colorado, Boulder, CO, 80309, USA}
\altaffiltext{6}{www.rcs2.org}
\

\begin{abstract} 
The Red-Sequence Cluster Surveys (RCS-1 and RCS-2) are large optical imaging
surveys optimized to create well-characterized catalogs of
clusters of galaxies up to a redshift of $\sim1$.
We describe our first cosmological analysis, using the self-calibration
technique, of a cluster sample
of $\sim1000$ from the 90 sq.~deg RCS-1, using optical 
richness as a mass proxy.
We obtain values for the cosmological parameters
$\Omega_m$ and $\sigma_8$ that are in excellent
agreement with the
year-three WMAP results.  Furthermore, the derived cluster richness-mass
relationship is entirely consistent with those measured directly using
dynamical mass measurements.

We describe briefly the on-going RCS-2, which is a 1000 sq.~deg survey
carried out using the one-square-degree MegaCam on the CFHT.
This survey, in  $z'$, $r$ and $g$ bands, will produce a cluster sample
of several 10$^4$ clusters up to $z\sim1$, and will allow
us to constrain the dark energy equation of state, $w$, and provide
a large cluster sample for the studies of cluster evolution and
lensing.
\end{abstract}

\section{Introduction}
Galaxy clusters, the largest mass concentrations in the
Universe, have long played significant roles in the study of
galaxy evolution and the determination of cosmological parameters.
In both cases, the leverage and advantages offered by having a
sample with a large redshift grasp are significant.
In the case of cluster evolution, at the redshift of $\sim1$ we are
approaching the formation epoch of clusters, often put at $z\sim2$ to 5, so
that evolutionary effects can be more easily discerned.
For the determination of cosmological parameters, a large baseline
of redshift allows us to measure the evolution of the mass function,
breaking the degeneracy between $\Omega_m$ and $\sigma_8$.
The measurement of the mass function at $z\sim1$ of a sufficiently
large sample will also provide a chance to constrain $w$.

The red-sequence method (Gladders \& Yee 2000) was developed with
the specific aim to create large, robust galaxy cluster samples efficiently
out to $z>1$ using optical data.
This technique uses the red-sequence (early-type) galaxies as markers
for clusters, searching for enhanced galaxy surface densities in 
successive cuts in the color-magnitude plane, which act
to provide a discrimination in the third dimension, mitigating
the projection contamination
problem inherent in searching for high redshift galaxy clusters.
This technique is inexpensive, as it requires
photometry from only two filters,
and can be used to map large regions of the sky efficiently.

In this paper 
we describe the application of the first Red-Sequence Cluster Survey (RCS-1)
to the determination of $\Omega_m$ and $\sigma_8$, and the on-going
1000 square degree RCS-2 survey, with the goal of measuring $w$.
The creation of these large cluster samples to $z\sim1$ also enables
us to carry out many other studies, some of which are presented
in these proceedings, including an analysis of the Butcher-Oemler
effect using over 1000 clusters (Ellingson et al., these proceedings),
and the discovery of a large number of strong gravitational lenses
(Gilbank et al., these proceedings).

\section{Measuring Cosmological Parameters with RCS-1}
The RCS-1 provides for the first time a cluster sample which is
large, well-characterized, and covering a sufficiently large
redshift range to measure $\Omega_m$ and $\sigma_8$ directly using
cluster abundance.  The details are reported in Gladders et al.~(2007);
we present a brief description of the results here.

The RCS-1 survey contains a total of $\sim90$ sq.~deg of imaging
data obtained using the CFHT-12K camera and the MOSAIC-II camera
on the CTIO 4m with coverage in the $R_c$ and $z'$ bands (see
Gladders \& Yee 2005 for details). 
Both cameras provide a field of view of $\sim$1/3 sq.~deg.
 We select a `best' sample using 72 sq.~deg based
on depth and photometric calibration accuracy.  The sample contains
956 clusters in the redshift range of $0.35<z<0.95$.  We use a somewhat
restricted redshift range which provides the most robust red-sequence
photo-$z$ and cluster richness estimates.  We use the cluster richness
parameter $B_{gcR}$ (see Gladders \& Yee 2005),
 computed using the net galaxy count in the
red-sequence, as a mass proxy.  Cluster richness, or other
similar measurements such as cluster light, has been shown
to be correlated with mass, $T_x$, and $L_x$ in a number of
investigations (e.g., Yee \& Ellingson 2003; Hansen et al.~2005).
We link the observable to mass ($M_{200}$) using a power-law
 $M_{200}= 10^{A_{Bgc}} B_{gcR}^\alpha (1+z)^\gamma$, where $\gamma$
allows for possible evolution in the relationship.
We use the self-calibration analysis technique suggested by \cite{mm03},
in which the parameters of the mass-observable relation
are simultaneously fitted with the cosmological parameters.

  The cosmological parameters are estimated by fitting the number
of clusters at different redshifts ($dN/dz (z)$)
to a mass limit $M_{lim}(z)$
to that expected from the Jenkins et al.~(2001) mass function,
using a Markov-Chain Monte-Carlo (MCMC) method.
We effectively fit two cosmological parameters ($\Omega_m$ and $\sigma_8$)
and three cluster parameters (the mass limit as a combination of the
$A_{Bgc}$ and $\alpha$ parameters, $\gamma$, and the fractional
scatter $f_{sc}$ in the mass-$B_{gcR}$ relation).  
We use WMAP priors for $h$ ($0.72\pm0.08$),
$\Omega_b$ (0.046, fixed), and $n$ (0.99, fixed).
The results are shown in Figure 1.
We obtain $\Omega_m=0.31^{+0.11}_{-0.10}$ and $\sigma_8=0.67^{+0.18}_{-0.13}$,
which are in excellent agreement with those from the year-3 WMAP
results \cite{wmap3}. 

\begin{figure}[!ht]
\plotfiddle{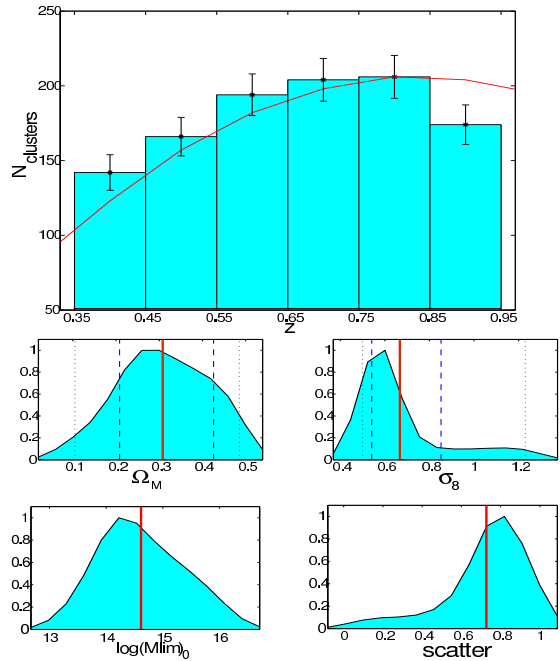}{8.5cm}{0}{90}{75}{-140}{-5}
\caption{Cosmological parameter analysis from the RCS-1.
Top Panel: Histogram of the observed cluster counts with
richness $B_{gcR}>300$. The solid line shows the best-fitting
cosmology + richness-mass model.
Central panels: The likelihood function of $\Omega_m$ and $\sigma_8$,
marginalized over the other fitting parameters.  The solid vertical
line in each panel indicates the mean value.
Bottom panels: The likelihood function of the fitted scatter and mass 
limit of the cluster sample. }
\end{figure}

An equally interesting result in our analysis is the comparison
between the cluster parameters derived from the self-calibration
cosmology fit to those measured explicitly from cluster samples,
specifically those obtained from the CNOC1 survey (Yee \& Ellingson 2003)
and a sample of 33 $z<0.6$ RCS-1 clusters with dynamical mass measurements
(Blindert et al. 2007).
We perform this comparison of the $B_{gc}$-mass relation by fitting
with 4 cluster parameters, replacing $M_{lim}$ with $A_{Bgc}$ and
$\alpha$.  
We find that the richness to mass relation as derived from the
cosmology fit produces entirely consistent results to those
from direct measurements of the relation. We obtain  $A_{Bgc}$
and $\alpha$ values of 10.55$^{+2.27}_{-1.71}$ and
$1.64\pm0.90$, respectively, compared to 
$9.89\pm0.89$ and $1.64\pm0.28$ 
using the CNOC1 sample (Yee \& Ellingson 2003).
Similar results are also obtained by Blindert et al.~(2007) from
the RCS-1 spectroscopic sample.
Furthermore, the best fit scatter $f_{sc}=0.73\pm0.22$ is also
in excellent agreement with that determined from the Blindert et al.
sample of $\sim0.7$.
The fact that
the fitting results for both the cosmological parameters and
the cluster parameters are in excellent agreement
with other completely independent measurements
is a strong endorsement of both the use of cluster mass function,
including that derived from an optical sample such as the RCS,
as a cosmological probe, and the  self-calibration methodology.

\section{The RCS-2}
The RCS-2 is the first large scale imaging survey with
sufficient area and depth for constraining $w$ using cluster abundance.
The recent commissioning of the MegaPrime Camera (MegaCam) on CFHT 
provides the field of view on a sufficiently large telescope required
to carry out such a survey for the first time.
The MegaCam is a one-square-degree camera consisting of a mosaic of
thirty-six 4612$\times$2048 pixels, with a scale of 0.18$"$/pix.
The survey will image 820 sq.~deg using $z'$, $r$, and $g$ filters
of relatively shallow depths with integration times of 6, 8 and
4 minutes, attaining $5\sigma$ magnitudes of 22.5, 24.8, and 25.3,
respectively.
A total coverage of $\sim$1000 sq.~deg is obtained  when
combined with the 170 sq.~deg of 
the wide component of the CFHT Legacy Survey.
The survey area for the 820 sq.~deg is divided up into 13 patches 
of 9$\times$9 or 6$\times$6 pointings of the MegaCam.
The survey is expected to be completed by the end of 2007.
The depths of the survey allow us to be able to detect galaxy
clusters of richness of about half that of Coma out to redshift
1.  The inclusion of the $g$ band filter will allow for the
proper characterization of the galaxy sample down to redshift as low
as 0.1.

In any survey of significant size, the only feasible way to obtain
the mass observable is from the survey data themselves, in our case, the
optical richness.
A large observational effort is also underway, using primarily
clusters from RCS-1, to calibrate and provide strong priors 
on the cluster mass-richness relation.
Having strong priors to the mass-richness relation over several
redshift bins (to estimate $\gamma$), and including an estimate
of the scatter, will provide much tighter constraints on the
cosmological parameters, and is essential for the measurement of $w$.
Our calibration efforts include using weak lensing cluster mass
as our primary mass calibrator, augmented by dynamical mass measurements
from multi-object spectroscopy
for a core sample of $\sim70$ clusters, and X-ray observations.
The study of cluster galaxy evolution, using multi-color optical
and IR photometry and spectroscopy, will also allow us to account
for galaxy evolution when computing the richness of clusters at different
redshifts.
The results from RCS-1 bode well for the use of RCS-2 as a probe
for $w$.
Scaling with the RCS-1 as a guide, we expect to obtain a cluster
sample of the order of 20,000 clusters with $0.15<z<1.0$ of similar
richness.  
With priors on the mass-richness relation from our follow-up program,
we expect to be able to measure to an accuracy of $\sim$10\%
in $w$, $\sim$0.02 in $\Omega_m$, and $\sim$0.05 in $\sigma_8$.



\end{document}